\documentclass[preprint,aps,amsmath,superscriptaddress,nofootinbib,tightenlines]{revtex4}
\usepackage{bm}
\usepackage{epsfig}

\newif\ifpdf
\ifx\pdfoutput\undefined
\pdffalse 
\else
\pdfoutput=1 
\pdftrue
\fi

\def\OMIT#1{}

\newcommand{\fr}{\frac} 
\newcommand{\ra}{\rightarrow}
\newcommand{\nn}{\nonumber}

\newcommand{\bea}{\begin{eqnarray}}
\newcommand{\eea}{\end{eqnarray}}

\newcommand{\tenb}{\bf{\overline{10}}}

\begin{document}

\ifpdf
\DeclareGraphicsExtensions{.pdf, .jpg}
\newcommand{\picspace}{\vspace{-2.5in}}
\newcommand{\picspacehalf}{\vspace{-1.75in}}
\else
\DeclareGraphicsExtensions{.eps, .jpg,.ps}
\newcommand{\picspace}{\vspace{0in}}
\newcommand{\picspacehalf}{\vspace{0in}}
\fi


\title{Excited $D_s$ (and Pentaquarks) in Chiral Perturbation Theory
\footnote{Presented at Cracow Epiphany Conference on Hadron Spectroscopy, 
6-8 January 2005, Cracow, Poland } 
}
\author{Thomas Mehen\footnote{Electronic address: mehen@phy.duke.edu}}
\affiliation{Department of Physics, Duke University, Durham,  NC 27708\vspace{0.2cm}}
\affiliation{Jefferson Laboratory, 12000 Jefferson Ave., Newport News, VA 23606\vspace{0.2cm}}

\date{\today\\ \vspace{1cm} }
\begin{abstract}
I present results of a heavy hadron chiral perturbation theory analysis of the
decays and masses of the recently discovered excited charm mesons. The
present data on the electromagnetic branching ratios are consistent with heavy
quark symmetry predictions and disfavor a molecular interpretation of these
states. I also discuss model independent predictions for the strong
decays of pentaquarks in the $\mathbf{\overline{10}}$ representation of $SU(3)$ which can 
be used to constrain the angular momentum and parity quantum numbers of these
states.
\end{abstract}

\maketitle
 
\section{Introduction} 

In recent years there has been a resurgence in hadron spectroscopy as many experiments have reported evidence for new hadrons. Examples include excited charm strange mesons
$D_{s0}(2317)$~\cite{Aubert:2003fg} and $D_{s1}(2460)$~\cite{Besson:2003cp}, their nonstrange partners~\cite{Anderson:1999wn,Abe:2003zm,Link:2003bd},  the exotic pentaquarks
$\Theta^+$~\cite{Hicks:2005gp},  $\Xi^{--}$~\cite{Alt:2003vb} and $\Theta_c(3099)$~\cite{Aktas:2004qf}, the new charmonium state
$X(3872)$~\cite{Choi:2003ue,Acosta:2003zx,Abazov:2004kp}  and doubly charm  baryons~\cite{Mattson:2002vu}. The status of these various hadrons varies greatly. For example,
the  $D_{s0}(2317)$ and $D_{s1}(2460)$ are firmly established~\cite{Eidelman:2004wy} while the existence of pentaquarks is quite controversial.

In this talk,  effective field theory methods are used to obtain model independent predictions for the properties of the excited charm mesons as well as pentaquarks.
These predictions yield qualitative insight into the nature of the novel states. Heavy hadron chiral perturbation theory (HH$\chi$PT)~\cite{w,bd,y},  which synthesizes
heavy quark symmetry for heavy mesons and the spontaneously broken chiral  symmetry which governs the low energy interactions of pions, can be used  to make predictions
for the electromagnetic and strong  decays of the $D_{s0}(2317)$ and $D_{s1}(2460)$. These predictions can be used to test the hypothesis that the $D_{s0}(2317)$ and
$D_{s1}(2460)$ are molecular bound states of $D K$ and $D^* K$,
respectively~\cite{Mehen:2004uj}.  I also discuss the puzzle of the $SU(3)$ splittings of the excited
states and attempts to address the problem within HH$\chi$PT~\cite{Becirevic:2004uv,Mehen:2005hc}. The successful prediction of parity doubling
models~\cite{Bardeen:2003kt,Bardeen:1993ae,Nowak:1992um,Nowak:2003ra} that the hyperfine splittings of the excited and ground state heavy meson doublets are equal is
shown to be robust at the one-loop level~\cite{Mehen:2005hc}. Finally, heavy baryon chiral perturbation theory~\cite{Jenkins:1990jv} is extended to include
pentaquarks and used to make parameter free predictions for certain ratios of two-body decays which constrain the angular momentum and parity quantum numbers of
the exotic states~\cite{Mehen:2004dy}.

\section{Electromagnetic and Strong Decays of $D_{s0}(2317)$ and $D_{s1}(2460)$}

The discovery of $D_{s0}(2317)$ \cite{Aubert:2003fg} and $D_{s1}(2460)$ \cite{Besson:2003cp} came as a surprise because quark
models~\cite{Godfrey:xj,Godfrey:wj} as well as lattice calculations~\cite{Hein:2000qu,Boyle:1997rk,Lewis:2000sv}  predicted  that these states would
lie above the threshold for kaon decays. If the $J^P=0^+$ and $J^P=1^+$ charmed strange mesons were above  this threshold, they would have been rather
broad resonances.  In fact, the $D_{s0}(2317)$ and $D_{s1}(2460)$ are about 40 MeV below the threshold for decay into $D K$ and $D^* K$, respectively.
The only kinematically allowed strong decays are  $D_{s0}(2317) \to D_s \pi^0$ and $D_{s1}(2460) \to D_s^* \pi^0$, which violate isospin. Therefore,
the states are quite narrow: $\Gamma[D_{s0}(2317)] < 4.6$ MeV and $\Gamma[D_{s1}(2460)] < 5.5$ MeV~\cite{Eidelman:2004wy}. Allowed electromagnetic
decays are
\bea 
 D_{s1}(2460) \to D_s^* \gamma \, ,\quad D_{s1}(2460) \to D_s  \gamma \, , \quad  D_{s0}(2317)\to D^*_s \gamma \, ,\nn
\eea
while the decay  $D_{s0}(2317)\to D_s \gamma$ is forbidden by angular momentum conservation. 

To date only the decay $D_{s1}(2460) \to D_s \gamma$ has been observed. 
Belle  has observed the decay $D_{s1}(2460) \to D_s\gamma$ from $D_{s1}(2460)$ produced in the decays
of $B$ mesons~\cite{Krokovny:2003zq} and from continuum $e^+e^-$ production~\cite{Abe:2003jk}.
The BaBar experiment has also recently reported observing this decay~\cite{Aubert:2004bp}.
The electromagnetic branching ratio obtained by averaging the three experimental measurements is shown in the first column of Table 1 
along with upper bounds on the unobserved electromagnetic branching ratios from the CLEO experiment~\cite{Besson:2003cp}.
(The Belle  collaboration  quotes weaker lower bounds for these ratios~\cite{Abe:2003jk}.)

\begin{table}[h!]
\begin{center} \begin{tabular}{ccc|cccccc}
&  && Expt. && Molecule && HQS &\\
\hline
&  &&  &&   &&   &\\
& {\Large $\frac{ \Gamma[D_{s1}(2460)\to D^*_s \gamma]}{\Gamma[D_{s1}(2460)\to D^*_s \pi^0]}$}
&& \quad$< \,0.16$ &&\quad  3.23 (1.08) &&\quad $0.32 \pm 0.05 \pm 0.10$ &\nn  \\
&  &&  &&   &&   &\\
& {\Large $\frac{ \Gamma[D_{s1}(2460)\to D_s \gamma]}{\Gamma[D_{s1}(2460) \to D^*_s \pi^0]}$}
&&\quad $0.39 \pm 0.06$ &&\quad  2.21 (0.74) &&\quad $0.39$ (fit)&\nn  \\
&  &&  &&   &&   &\\
& {\Large $\frac{\Gamma[D_{s0}(2317)\to D_s \gamma]}{\Gamma[D_{s0}(2317) \to D_s \pi^0]}$}
&& \quad$< \,0.059$ &&\quad  2.96 (0.99)&&\quad $0.12 \pm 0.02 \pm 0.04$ &\nn  
\end{tabular}
\end{center}
\caption{Electromagnetic Branching Fraction Ratios}
\end{table}

The low mass of the $D_{s0}(2317)$ and $D_{s1}(2460)$ has prompted speculation that these states are exotic. Possibilities include $D K$
molecules~\cite{Barnes:2003dj,Nussinov:2003uj,Chen:2004dy}, $D_s \pi$ molecules~\cite{Szczepaniak:2003vy}, and tetraquarks
\cite{Nussinov:2003uj,Cheng:2003kg,Browder:2003fk,Terasaki:2003qa,Terasaki:2003dj,Terasaki:2004yx, Vijande:2003wk}. The proposal  that these are $D K$ molecules, in addition to
resolving the discrepancy with model predictions of the masses, could potentially explain why the hyperfine splitting between the $D_{s0}(2317)$ and $D_{s1}(2460)$
is equal to the hyperfine splitting of the ground state $D$ meson doublet to within a few MeV. This hypothesis can be tested using chiral perturbation theory. 

If the $D_{s0}(2317)$ is a molecular bound state of $D$ and $K$, then the typical three-momentum of its constituents is $p = \sqrt{2 \mu B} \approx 190$ MeV, where $\mu$ is
the reduced mass and $B$ is the binding energy. This means that both constituents are nonrelativistic. Corrections to the nonrelativistic approximation are $\sim p^2/m_K^2
\sim 0.16$. For a nonrelativistic  bound state, the decay rate  can be expressed as a product of the wavefunction at the origin and a transition matrix element involving its
constituents. For example, if the $D_{s0}(2317)$ is a bound state of $D$ and $K$, then the electromagnetic decay amplitude for the $D_{s0}(2317)$ is 
\bea
{\cal M}[D_{s0}(2317) \to D_s^* \gamma] &\propto& \int d^3 \vec{p} \, |\tilde \psi(\vec{p} \,)|^2 
{\cal M}[D(\vec{p} \,) K(-\vec{p} \,) \to D^*_s \gamma] \nn \\
&\propto& |\psi(0)|^2 {\cal M}[D K\to D^*_s \gamma] \, .
\nonumber
\eea
Here $\tilde \psi(\vec{p}\,)$ is the momentum space wavefunction and $\psi(0)$ is the position space wavefunction at the
origin. In the last line the matrix element ${\cal M}[D K\to D^*_s \gamma]$ has been expanded to lowest order in $p$. To calculate 
$\psi(0)$ requires detailed knowledge of the mechanism that binds the $D K$ into a composite hadron. Such a calculation
is necessarily nonperturbative. However, this factor cancels out of the ratios in Table~1.
The experimentally observed branching ratios are then determined by ratios of the amplitudes for $D^{(*)}K \to D_s^{(*)}\gamma$ 
and $D^{(*)}K \to D_s^{(*)}\pi^0$  at threshold. These were computed using HH$\chi$PT in Ref.~\cite{Mehen:2004uj}.  
\begin{figure}[!t]
  \centerline{\epsfysize=7.5truecm \epsfbox[90 490 480 730]{DK_gamma.ps}  }
 {\caption[1]{Leading order diagrams for $D^{(*)}K$ bound states decaying 
into $D_s^{(*)} \gamma$. The shaded oval represents the $D^{(*)} K$ 
bound state wavefunction. }
\label{DK_photo} }
\end{figure}
\begin{figure}[!t]
  \centerline{\epsfysize=3.75truecm \epsfbox[0 600 550 720]{DK_pion.ps}}
 {\caption[1]{Leading order diagram for $D^{(*)}K$ bound states decaying 
into $D_s^{(*)} \pi^0$. 
The dashed line from the  bound state is a $K$, the dashed line 
in the final state is an $\eta$ which mixes into a $\pi^0$. }
\label{DK_strong} }
\end{figure}

The diagrams for electromagnetic and strong decays are shown in Figs. \ref{DK_photo} and~\ref{DK_strong}, respectively. The diagrams in Figs.2b and 2c only  contribute to the
P-wave channel so the entire contribution to the strong decay comes from the graph in Fig.~2a.  Dashed lines are Goldstone bosons, wavy lines are photons and the double
lines are heavy mesons. The blob represents the bound state wavefunction   and the cross represents the isospin violating  $\pi^0-\eta$ mixing term. The coupling of heavy
mesons and Goldstone bosons to photons comes from gauging the kinetic terms in the HH$\chi$PT Lagrangian and the coupling of the heavy mesons to Goldstone bosons is
proportional to the axial coupling, $g$, of the heavy mesons. This coupling is known from the strong decay of the $D^*$. At this order the molecular scenario
makes predictions for the electromagnetic branching fraction ratios, which are shown in the column labeled ``Molecule" in Table 1. The results depend on two  parameters: $g$
and the Goldstone boson decay constant, $f$.  In this calculation $g=0.27$~\cite{Stewart:1998ke}.  At lowest order, $f=f_\pi=f_K=f_\eta$ but $SU(3)$
breaking leads to different decay constants for pions, kaons and etas.  In the calculation of the electromagnetic decays, $f=f_K=159$ MeV is used since these decays involve
kaons only. Two different values of $f$ are used in the calculation of the strong decays. The first number in the second column of Table~1 corresponds to using
$f=f_\eta=171$ MeV  and the number in the parenthesis corresponds to using $f=f_\pi=130$ MeV. The difference gives a crude estimate of the uncertainty due to  higher order
$SU(3)$ breaking effects. Because the matrix element squared for the strong decay is $\sim f^{-4}$, the magnitude of the branching fraction ratios is highly uncertain.
However, even allowing for this considerable uncertainty, the electromagnetic branching fraction ratios are badly overpredicted in the molecular scenario. The branching
fraction ratios are proportional to $g^2$, so larger values of $g$ which are sometimes used in the literature will lead to even larger disagreement with experiment. Also,
the relative sizes of the branching ratios is qualitatively incorrect. The ratio $\Gamma[D_{s1}(2460)\to D_s \gamma]/\Gamma[D_{s1}(2460)\to D_s^* \pi^0]$ is predicted to be
the smallest rather than  the largest as is experimentally observed. The molecular hypothesis is in disagreement with the data on electromagnetic decays.

An alternative approach is to use heavy quark symmetry to relate the electromagnetic decays and strong decays. At the level of HH$\chi$PT 
this is implemented by adding the excited $J^P=0^+$ and $J^P=1^+$ states to the Lagrangian by hand in a manner consistent with heavy quark symmetry~\cite{Falk:1992cx}.
A single operator in the HH$\chi$PT Lagrangian mediates all three electromagnetic decays and another operator mediates the two strong decays of the excited 
$D$ mesons, so the electromagnetic branching ratios can be predicted in terms of a single parameter which is fit to 
the observed value of $\Gamma[D_{s1}(2460)\to D_s \gamma]/\Gamma[D_{s1}(2460) \to D^*_s \pi^0]$~\cite{Mehen:2004uj}. 
The heavy quark symmetry prediction appears in the column labeled ``HQS" in Table 1. 
The other two ratios can then be predicted. 
The first error is due to experimental uncertainty in  $\Gamma[D_{s1}(2460)\to D_s \gamma]/\Gamma[D_{s1}(2460) \to D^*_s \pi^0]$, the second 
error is a 30\% uncertainty due to $O(\Lambda_{\rm QCD}/m_c)$ corrections to heavy quark symmetry. The experimental upper bounds on the 
unobserved branching ratios are below the predicted 
central values but are within expected errors.

\section{Charmed Meson Masses in HH$\chi$PT}

Experiments also claim to observe the nonstrange partners of the $D_{s0}(2317)$ and $D_{s1}(2460)$~\cite{Anderson:1999wn,Abe:2003zm,Link:2003bd}.  The $J^P=0^+$ and
$J^P=1^+$ nonstrange charm mesons are above the threshold for isospin conserving strong decays into $D \pi$ which makes these states much broader than their nonstrange
counterparts. The experimental average for the mass of the $D_0^0(J^P=0^+)$ is 2308 $\pm$ 36 MeV, and the mass of the $D_1^0(J^P=1^+)$ is 2438 $\pm$ 31 MeV. The $SU(3)$
splitting of the excited charm mesons is 9 $\pm$ 36 MeV for the $J^P=0^+$ mesons and 21 $\pm$ 31 MeV for the $J^P=1^+$ mesons. This is surprising  because typically
$SU(3)$ splittings between strange and nonstrange particles is $\sim 100$ MeV. 

Ref.~\cite{Becirevic:2004uv} made the first attempt to address this problem within the framework of HH$\chi$PT. These authors calculated
$\Delta m_{u/d} - \Delta m_s$, where $\Delta m_{u/d}$ is the splitting between the spin-averaged mass of the even-parity  
and odd-parity heavy meson doublets in the nonstrange sector, while $\Delta m_s$ is the analogous quantity in the strange
sector. Numerically, $\Delta m_s = 348$ MeV while $\Delta m_{u/d} \approx 430$ MeV. The calculation of Ref.~\cite{Becirevic:2004uv} works 
in the heavy quark limit so all $\sim 1/m_c$ suppressed operators are neglected. A linear  
combination of $SU(3)$ breaking counterterms contributing to $\Delta m_{u/d} - \Delta m_s$ is fixed from lattice calculations
of the quark mass dependence of $\Delta m_{u/d}-\Delta m_s$. Ref.~\cite{Becirevic:2004uv} then finds  $\Delta m_{u/d} - \Delta m_s \approx
- 100$ MeV, which has the wrong sign!

Recently, Ref.~\cite{Mehen:2005hc} improved upon the calculation of Ref.~\cite{Becirevic:2004uv} by systematically including all $O(1/m_c)$  and $SU(3)$ breaking
counterterms. Unfortunately this leads to a large number of free parameters appearing in the one loop calculation. These were determined by fitting to the observed spectrum.
Two fits were performed in Ref.~\cite{Mehen:2005hc}. The first used the value of $g$ extracted from Ref.~\cite{Stewart:1998ke}. Another axial coupling, $h$, was extracted
from a  tree level fit to the widths of the excited nonstrange charm mesons~\cite{Mehen:2004uj}. For these values of $g$ and $h$, the fit systematically underpredicts the
excited nonstrange meson masses, similar to the result of Ref.~\cite{Becirevic:2004uv}. However, the extractions of $g$ and $h$ use calculations that make different
approximations than are used in the one loop mass calculations. Therefore, those values of $g$ and $h$ may not be the correct parameters for the mass calculation. In the
second fit of Ref.~\cite{Mehen:2005hc}, the couplings $g$ and $h$ were treated as free parameters. This fit is highly underconstrained and it is possible to find regions of
parameter space where the observed spectrum can be reproduced. However, in these fits some of the $O(1/m_c)$ suppressed operators have uncomfortably large coefficients. The
$SU(3)$ splitting of the excited states remains puzzling in HH$\chi$PT.  An alternative approach is to extend the quark model to include couplings to the $DK$ continuum and
try to explain the spectrum via the coupled channel effect~\cite{Hwang:2004cd,Lee:2004gt,Simonov:2004ar,vanBeveren:2003kd,vanBeveren:2003jv,vanBeveren:2004ve}.

Parity doubling models~\cite{Bardeen:2003kt,Bardeen:1993ae,Nowak:1992um,Nowak:2003ra} of heavy hadrons make a tree level prediction that the axial couplings and
hyperfine splittings of the even- and odd-parity doublets are equal. The prediction for the hyperfine splittings of the charm strange mesons  is in good agreement with
data while the uncertainties  in the masses of the nonstrange excited charm mesons are  too large to test this prediction. The analysis of Ref.~\cite{Mehen:2005hc}
reveals that the region of HH$\chi$PT parameter space predicted by the parity doubling  model  is invariant under the renormalization group flow of HH$\chi$PT at one
loop.  It is encouraging to see that the parity doubling predictions are robust at the one loop level. Data is currently not accurate enough to test parity doubling
model predictions for axial couplings~\cite{Mehen:2004uj,Mehen:2005hc}.

\section{Strong Pentaquark Decays}

In this section I briefly describe the results of Ref.~\cite{Mehen:2004dy}
which applied heavy baryon chiral perturbation theory~\cite{Jenkins:1990jv} to the strong decays
of exotic pentaquarks in the $\mathbf{\overline{10}}$ representation. The exotic pentaquarks in the
$\mathbf{\overline{10}}$ are the $\Theta^+$, $\Xi^{--}$ and $\Xi^+$. Various experiments have reported
evidence for the $\Theta^+$ while evidence for the exotic cascades comes from the 
NA49 experiment~\cite{Alt:2003vb}. There are also  several experiments that do not see the pentaquarks~\cite{Hicks:2005gp}. 
The allowed two-body strong decays for these states are 
\bea
\Theta^+ \to p K^0, n K^+ \quad\, \Xi^{--} \to \Xi^- \pi^-, \Sigma^- K^-
\quad\, \Xi^{+} \to \Xi^0 \pi^+, \Sigma^+ \overline{K}^0 \, . \nn
\eea
All these decays are related by $SU(3)$. Only two-body decays are  kinematically  
allowed for the $\Theta^+$ while  the $\Xi$'s should also have multi-body decays.
There are experiments claiming to see the $\Theta^+$ in both $p K^0$ and $n K^+$ channels.
The NA49 experiments has only seen the $\Xi^{--}$ state in the $\Xi^- \pi^-$ channel.

Of course the primary experimental problem regarding the $\Theta^+$ and exotic $\Xi$'s  is firmly establishing whether or not these states actually
exist. If the pentaquarks are confirmed,  the most important experimental problem will be to determine their angular momentum and parity quantum numbers, $J^P$, which can
distinguish between various pentaquark models. The most commonly discussed  method is to measure  production of the $\Theta^+$ in polarized $pp$
collisions near the  production threshold~\cite{Thomas:2003ak,Hanhart:2004re}. Currently, experimental data on this process is unavailable. 

The main point of Ref.~\cite{Mehen:2004dy} is that interesting constraints on $J^P$ can be obtained by measuring two-body decays
of the exotic members of the $\tenb$ multiplet. The $J^P$ quantum numbers of the 
pentaquark determine the angular momentum, $L$, of the pion or kaon emitted in the decay. The rates are proportional 
to an $SU(3)$ Clebsch-Gordan coefficient times a phase space factor,
\bea
p^{2L+1} \quad (L \neq 0) \, , \qquad E^2 p\quad (L=0) \, ,\nn
\eea
where $p$ is the three-momentum and $E$ is the energy of the pion or kaon emitted in the decay. Therefore, the ratio  
$\Gamma[\Xi^{--} \to \Sigma^- K^-]/\Gamma[\Xi^{--} \to \Xi^- \pi^-]$, for example, is determined
entirely by $SU(3)$ and kinematic factors.   Results
for some interesting ratios are given in Table 2. Lower bounds on the ratio
of the total widths of  two exotic pentaquarks are also shown. These are lower bounds because the $\Xi$'s can have multi-body 
decay modes while the width of the $\Theta^+$ is saturated by two-body decays.
 The errors quoted are $30$\%, which is the
typical size of $SU(3)$ breaking. The ratios $\Gamma[\Xi^{--} \to \Sigma^- K^-]/\Gamma[\Xi^{--} \to \Xi^- \pi^-]$
and $\Gamma[\Xi^{0} \to \Sigma^+ K^-]/\Gamma[\Xi^{0} \to \Xi^- \pi^+]$ can discriminate between $J^P= \frac{1}{2}^-$
and $J^P= \frac{1}{2}^+$, which are the most common quantum number assignments that appear in existing pentaquark models.
If one finds that $\Gamma[\Xi^{--}], \Gamma[\Xi^+] < 10 \, \Gamma[\Theta^+]$, then $J^P=\frac{3}{2}^-$ and $J \geq \frac{5}{2}$ can
be ruled out.

\begin{table}[t]\label{penta}
\begin{center} \begin{tabular}{cccc}
 & \multicolumn{3}{c}{$J^P$} \\
\cline{2-4}
 & $ \qquad  \fr{1}{2}^- \qquad $ & $ \qquad  \fr{1}{2}^+ $  , \  $ \fr{3}{2}^+ \qquad$ & $  \fr{3}{2}^- $ \\[2mm]
 \hline  & &  & \\[-2mm]
 {\large  $ \fr{\Gamma(\Xi^{--}_{\tenb} \ra \Xi^- \pi^-)}{\Gamma(\Xi^{--}_{\tenb} \ra \Sigma^- K^-)} $} 
 & $1.2 \pm 0.4$  & $ 3.1 \pm 0.9$ &  $4.7  \pm 1.4$ \\[4mm]
{\large $ \fr{\Gamma(\Xi^{0}_{\tenb} \ra \Xi^- \pi^+)}{\Gamma(\Xi^{0}_{\tenb} \ra \Sigma^+ K^-)} $} 
 & $1.1 \pm 0.3$ &$ 2.9 \pm 0.9$ & $4.2 \pm 1.3$  \\[4mm]
\hline& & &  \\[-2mm]
{\large $ \fr{\Gamma(\Xi^{--})}
             {\Gamma(\Theta^+ ) } $}
 & $> 1.8 \pm 0.5$   & $ >  5.3 \pm 1.6 $ &$ >  14. \pm 4.$  \\[4mm]
\end{tabular}
\caption{Exotic Pentaquark Decay Ratios for various $J^P$}
\end{center} \end{table}

\section{Conclusion}

In this talk I described applications of chiral perturbation theory  to the strong interactions 
of newly discovered hadrons. HH$\chi$PT was applied to the electromagnetic and strong decays of the $D_{s0}(2317)$ and $D_{s1}(2460)$. 
Existing data is consistent with heavy quark symmetry predictions
and is inconsistent with a molecular interpretation of these states. The $SU(3)$ splitting of the
excited even-parity charm mesons  is puzzling.  The one-loop HH$\chi$PT formulae for the mass spectrum
contains a large number of free parameters from $1/m_c$ operators and axial couplings that are not well
determined, so it is not possible to make predictions for the spectrum. 

Parity doubling models make the prediction that the axial couplings and hyperfine splittings of the even-parity  and odd-parity heavy mesons are equal. This was shown to
hold at one loop order. The hyperfine splittings of the charm strange mesons are in agreement with this prediction. Currently data is not  accurate enough to seriously
constrain the axial couplings of the excited states. It would be interesting to obtain lattice calculations of these  couplings  to test the
parity doubling scenario as well as reduce theoretical uncertainty in HH$\chi$PT calculations. It would also be interesting to observe
the even-parity excited bottom strange mesons, who are also also predicted to lie below the kaon decay threshold~\cite{Mehen:2005hc} and should therefore be quite narrow. 

Finally, I discussed $SU(3)$ predictions for the strong decays of exotic pentaquarks and showed how these
can be used to constrain their $J^P$ quantum numbers.

\section{Acknowledgements}

This research has been supported by DOE grants DE-FG02-96ER40945 and DE-AC05-84ER40150. I thank the Institute
for Nuclear Theory at the University of Washington for its hospitality during the completion of this work.
It is a pleasure to acknowledge collaboration with R. Springer and C. Schat.

\end{document}